\documentclass[conference]{IEEEtran}
\IEEEoverridecommandlockouts

\usepackage{cite}
\usepackage{amsmath,amssymb,amsfonts}
\usepackage{algorithmic}
\usepackage{graphicx}
\usepackage{textcomp}
\usepackage{xcolor}

\usepackage[utf8]{inputenc}	
\usepackage[T1]{fontenc}	
\usepackage{CJKutf8}	

\usepackage{makecell}	
\usepackage{multirow}	
\usepackage[numbers]{natbib}	
\usepackage{orcidlink}	
\usepackage{subcaption}	
\usepackage{tabularx}	

\def\BibTeX{{\rm B\kern-.05em{\sc i\kern-.025em b}\kern-.08em
    T\kern-.1667em\lower.7ex\hbox{E}\kern-.125emX}}

\begin{document}
\begin{NoHyper} 

\begin{CJK}{UTF8}{gbsn}	

%

\title{How to check in continually over 4,000 days on an online learning platform? An empirical experience and a practical solution\\
\thanks{This work is funded by the Research Initiation Fund Project of Guangzhou Institute of Science and Technology (2023KYQ184).}
}

\author{\IEEEauthorblockN{Jialiang Lin}
\IEEEauthorblockA{\textit{School of Computer Science and Engineering, Guangzhou Institute of Science and Technology}\\
Guangzhou, China \\
me@linjialiang.net \\
https://orcid.org/0000-0001-8343-0230}
}

\maketitle

\begin{abstract}
The check-in service is often provided as an incentive system by online learning platforms to help users establish a learning routine and achieve accomplishment. However, according to the questionnaire conducted in this study, 82.5\% of users of online English learning platforms that feature a check-in service have failed to maintain the daily check-in behavior for long-term language learning, mainly by reason of demotivation, forgetfulness, boredom, and insufficient time. As a language learner, I have an empirical experience in maintaining a record of over 4,000 daily check-ins on China's leading online English learning platform of Shanbay. In the meantime, I have been constantly exploring a practical solution to help cultivate perseverance for other users to follow through the learning routine. In this paper, I systematically introduce this practical solution, the GILT method, and its instructions. The experience and solution for perseverance development are based on Shanbay, but they can be applied to other learning platforms for different purposes.
\end{abstract}


\section{Introduction}
\label{sec:introduction}

Perseverance is one of the two key factors of grit that contribute to success over a long period of time~\citep{duckworth-grit-2007}, and grit is positively correlated with achievements in many fields~\citep{duckworth-grit-2016}. Compared to the other key factor of passion, perseverance is a stronger predictor of achievement~\citep{crede-much-2017}. Perseverance is particularly predictive in the field of education, especially for second language acquisition (SLA)~\citep{dornyei-innovations-2020}. SLA is a long-term undertaking, which is measured in years or even decades. To get into a sustained commitment for such a long period no doubt demands perseverance~\citep{khajavy-closer-2021}.

Recent years have seen dramatic growth in the integration of information technology and education, leading to a transformative evolution in teaching and learning methodologies~\citep{godsk-engaging-2024}. The check-in service, extensively adopted by online learning platforms, is designed to directly assist users in cultivating perseverance~\citep{nie-using-2020,zhang-does-2024}. It is similar to the system of tracking students' attendance to foster discipline and achievement. Users of online learning platforms can check in on the platform after completing assigned learning tasks. The platforms will automatically record data such as the content studied, the duration, and the consecutive check-in dates, and visualize the data through charts. The check-in service can also serve as an interactional feature as it is also optional for users to share their check-in records on social media.

The check-in service is helpful in general. Nonetheless, it is not effective for every online learning platform user in achieving perseverance~\citep{jin-users-2018}. It is still challenging for a large number of online learning platform users to engage in language learning on a regular basis. The questionnaire conducted in this study (described in detail in Section~\ref{sec:reasons}) reveals that fewer than 20\% of online English learning platform users are able to persevere in daily check-in for language learning. Many factors can result in user abandonment of the learning routine and the failure of the check-in system. The benefits of the check-in service are straightforward, but to cultivate actual perseverance in language learning requires additional effort.

I began using one of the most popular and representative online English learning platforms, Shanbay, on May 17th, 2013. This platform also features the check-in service, and I have maintained a continual check-in record for over 4,000 days (Figure~\ref{fig:snapshot-4000}). In this paper, I share my empirical experience and propose a practical solution to help users persevere in language learning through online platforms.

\begin{figure}[htb]
	\centering
	\includegraphics[width=0.42\textwidth]{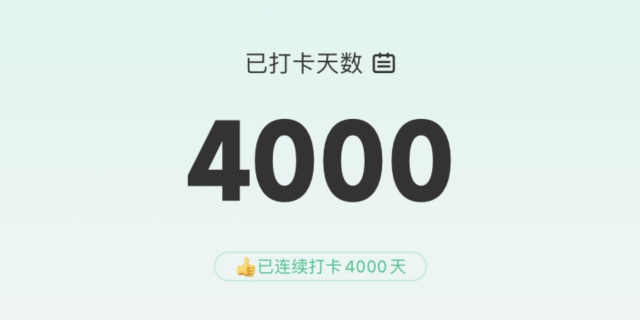}
	\caption{Screenshot of the author's 4,000-day continual check-in record on Shanbay, taken on April 28th, 2024. The top line in Chinese displays ``Days checked in'', while the bottom line in Chinese indicates ``Continually checking in for 4,000 days''.}
	\label{fig:snapshot-4000}
\end{figure}

\section{Reasons of User Abandonment}
\label{sec:reasons}

\subsection{Questionnaire}

To gain insight into the check-in behavior on online English learning platforms, a questionnaire is designed and distributed through the WJX questionnaire platform. Invalid responses with inconsistent answers are screened out. Altogether, 389 valid responses to the questionnaire are collected. The basic information of these respondents is shown in Table~\ref{tab:basic-info-respondent}.

\begin{table}[htbp]
\centering
\caption{Basic Information of Respondents}
\label{tab:basic-info-respondent}

\begin{tabular}{l l r r}
\hline
\textbf{Demographic} & \textbf{Option} & \textbf{Frequency} & \textbf{Percentage} \\
\hline
\multirow{6}{*}{\textbf{Age}}  
 & 0--9 & 11 & 2.8 \\
 & 10--19 & 78 & 20.1 \\
 & 20--29 & 179 & 46.0 \\
 & 30--39 & 88 & 22.6 \\
 & 40--49 & 26 & 6.7 \\
 & 50+ & 7 & 1.8 \\
\hline
\multirow{3}{*}{\textbf{Gender}}  
 & Female & 224 & 57.6 \\
 & Male & 163 & 41.9 \\
 & Non-binary & 2 & 0.5 \\
\hline
\multirow{3}{*}{\textbf{\makecell[l]{Educational \\ background}}}  
 & Below undergraduate & 61 & 15.7 \\    
 & Undergraduate & 234 & 60.1 \\    
 & Postgraduate & 94 & 24.2 \\    
\hline
\multirow{4}{*}{\textbf{\makecell[l]{Employment \\ status}}}  
 & Student & 251 & 64.5 \\
 & Employed & 129 & 33.2 \\
 & Unemployed & 7 & 1.8 \\
 & Retired & 2 & 0.5 \\
\hline
\textbf{\makecell[l]{Total \\ responses}} & & 389 & 100 \\
\hline
\end{tabular}
\end{table}

Among these valid respondents, 80.7\% claim that they have used online English learning platforms that feature a check-in service. Of these users, 82.5\% report that they have stopped checking in at some point or have fully withdrawn from the platforms. A follow-up question is asked to these respondents: what are the main reasons that make you stop checking in temporarily or abandon the platform entirely? This is a multiple-choice question, and respondents are required to select at least one option. The order of the options is randomized in order to minimize the selection bias. Of the 259 respondents, 134 check the box of ``laziness or unwillingness''. This is related to learning attitude, which is a broad topic in the field of education and beyond the scope of this paper. Therefore, this option is excluded from further analysis and the results are presented in Figure~\ref{fig:abandonment-reasons}.

\begin{figure*}[htb]
	\centering
	\includegraphics[width=0.65\textwidth]{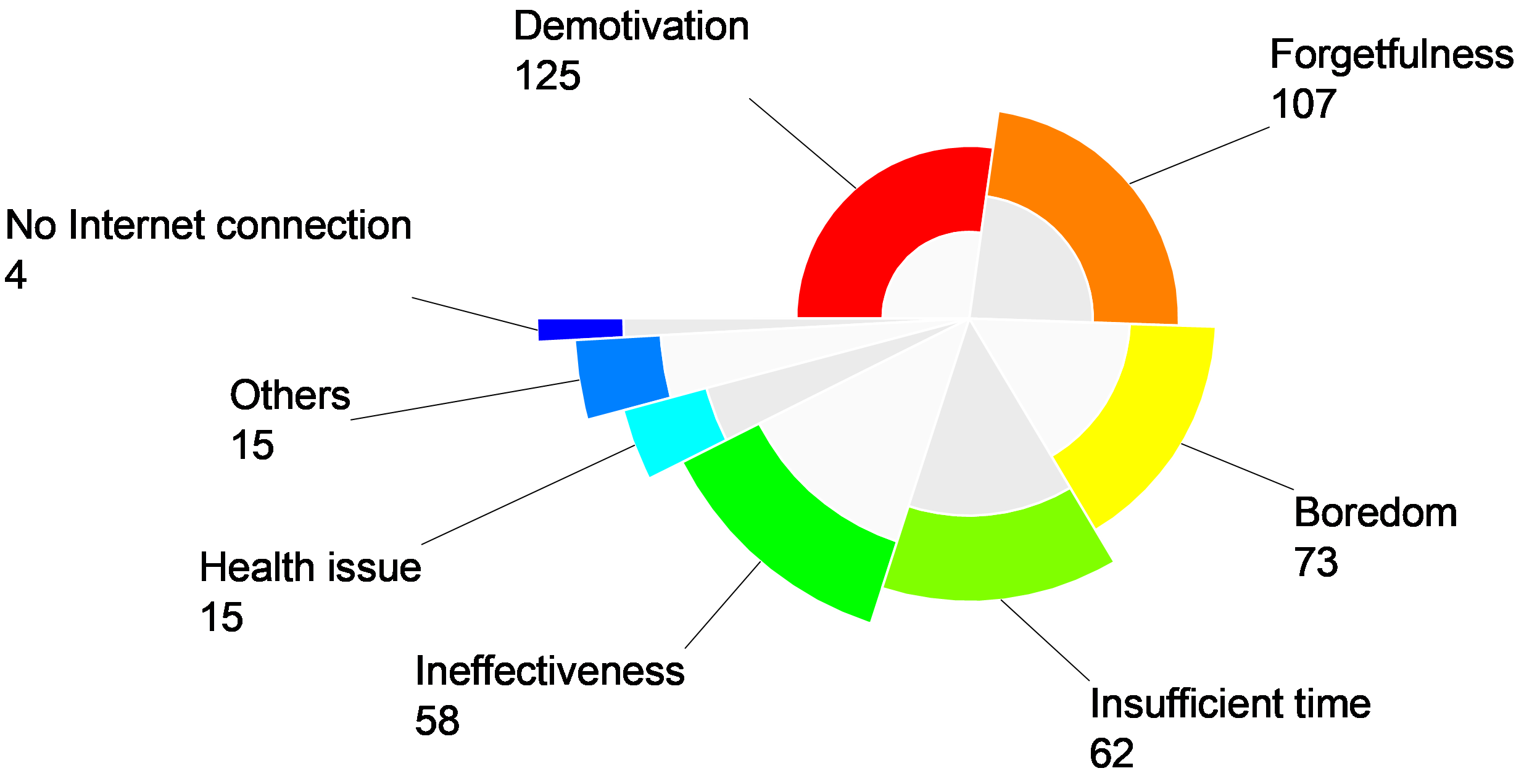}
	\caption{Reasons for abandoning online English learning platforms with a check-in service}
	\label{fig:abandonment-reasons}
\end{figure*}

As revealed in Figure~\ref{fig:abandonment-reasons}, demotivation, forgetfulness, boredom, and insufficient time are the top four reasons for user abandonment of online English learning platforms with a check-in service.

\subsection{Analysis}

\textbf{Demotivation.} Motivation, whether explicit or implicit, is the driving force behind people's actions, especially those long-term ones~\citep{heckhausen-motivation-2018}. The eagerness to achieve goals determines their motivation for and success in long-term learning. Gardner's socio-educational model is a classical theoretical framework that classifies SLA motivation into intrinsic motivation and extrinsic motivation~\citep{gardner-motivation-2010}. Originating from within individuals, intrinsic motivation comes from the appreciation of the learning process where individuals accomplish a specific learning task through self-improvement. In contrast, extrinsic motivation is motivation from outside of individuals, including external rewards and punishments associated with success or failure in a specific learning task~\citep{walker-identification-2006}. The intrinsic motivation for SLA can be attributed to the strong interest in or identification with the target language, while extrinsic motivation for SLA is typically associated with instrumental goals such as passing exams, earning qualifications, or securing job promotions. However, it is not easy for SLA beginners to develop such strong interest or identification. Many people start using Shanbay with extrinsic motivation. However, learning outcomes are more strongly correlated with intrinsic motivation than with extrinsic motivation~\citep{lei-intrinsic-2010}. Moreover, once the extrinsic need is satisfied, for example, after passing an exam, people often suffer from a sudden motivation loss in the action that is originally driven by this need. This is an important reason for user abandonment of online English learning platforms.

\textbf{Forgetfulness.} When a task is completed every day, it will gradually become a fixed part of a person's daily routine~\citep{keller-habit-2021}. The daily routine is a sequence of tasks regularly being conducted at specific times. Each task has its secure time slot in a day, for example, going for a run after waking up or reading the news before work. Ideally, this daily routine is to be maintained if there is no interruption of special events. However, such disruptions are inevitable in most cases. This is the same with the check-in task on Shanbay for English learning. After it gains its time slot in a user's daily routine, special events, such as meetings or accidents, can still interfere and take over that assigned time slot. When this happens and with the following task waiting to be done at the next time slot, the user can easily forget the original check-in task. This is a common issue reported by users on the Shanbay community and other social media, where they completely forget to finish the daily task of English learning and checking in. Worse still, if the assigned time slot is repeatedly being occupied by unexpected special events, the original task will lose its position in the daily routine fully. This situation is especially common during the initial phase of routine and habit formation.

\textbf{Boredom.} Shanbay offers creative learning content in vocabulary, listening, speaking, and reading activities alongside the learning and check-in services. These features are more appealing to new users of the platform. At the beginning of their Shanbay experience, users are more likely to have stronger interest, passion, and engagement. However, as time goes by, the content and services of the platform will naturally become less intriguing, leading to a decline in user passion. This phenomenon is known in psychology as ``habituation''~\citep{thompson-habituation-2009}, defined as ``a behavioral response decrement that results from repeated stimulation and that does not involve sensory adaptation/sensory fatigue or motor fatigue''~\citep{rankin-habituation-2009}. \citet{kim-temporal-2009} reveal that passion fades significantly after 6 months, accompanied by changes in brain activation patterns, a result of normal psychological processes. As individuals start to lose interest after this period, the feeling of boredom manifests as a state of reduced arousal and dissatisfaction caused by a lack of sufficient stimulation. Accumulated feelings of boredom over time can cause learning fatigue and eventually result in user abandonment.

\textbf{Insufficient time.} Everyone has only 24 hours in a day, which is one of the fairest aspects in the world, and that is why time management is a crucial skill. Managing online check-in for English learning amid other daily tasks depends on whether adequate attention is assigned to this task for prioritization over others. The Eisenhower Matrix is one of the most effective and popular time management methods for prioritizing actions throughout a day~\citep{kirillov-theory-2015}. With two intersecting axes of importance and urgency, the Matrix is divided into four quadrants: ``important and urgent'', ``important but not urgent'', ``not important but urgent'', and ``neither important nor urgent''. Based on this framework, daily tasks can also be categorized accordingly. Important tasks, whether urgent or not, are those truly worthy of our time and attention. For tasks of both importance and urgency, they should be addressed immediately. For tasks of urgency but not importance, they ought to be handled as efficiently as possible, minimizing the time spent on them. The daily check-in for long-term language learning falls into the category of being important but not urgent and should be assigned with a proper time slot to ensure completion. However, individuals who lack sufficient self-discipline often find themselves preoccupied with tasks of urgency but not importance, leaving no time for tasks of importance but not urgency. This situation is prone to occur during the initial phase of using Shanbay, when checking in on the platform has not yet developed into a habit for the user to carry out naturally.

\section{The GILT Method}

\label{sec:methods}

As a loyal user with a continual check-in record of over 4,000 days on Shanbay, I have been recommending the platform to many English learners in both public and private settings for more than a decade. Drawing on my first-hand experience and diverse user feedback, I propose the GILT method as the practical and feasible solution to address user abandonment of online learning platforms that feature a check-in service. GILT stands for goal setting, incentive planning, light starting, and team learning.

\subsection{Goal Setting}
\label{subsec:goal}

Goals can define one's ultimate performance. \citet{custers-unconscious-2010} suggest that human behavior is quite often guided by goals, both consciously and unconsciously. Goals can even facilitate the formation of certain behaviors at an unconscious level by directing attention toward things related to the goal~\citep{dijksterhuis-goals-2010}. When people feel like giving up, the desire to reach the goal helps them persevere. According to the goal-setting theory~\citep{locke-building-2002}, appropriate goals are able to strengthen students' motivation in learning. Therefore, setting a proper goal is the key to long-term language learning. A proper goal should be specific, difficult, and attainable. First, a goal should be defined specifically to serve as a precise guide to the goal pursuer, reducing confusion and misunderstanding. This is the basis for effective planning, execution, and evaluation. Many Shanbay users set ``improving English'' as the goal. While well-intentioned, this goal is too vague to be effective, because it is not clear and cannot be measured. There are no unified criteria for language improvement, and the users will not be able to decide whether the goal is achieved or not, thus negating the purpose of goal setting. Second, a proper goal should pose a moderate level of difficulty to the learner. Achieving goals without effort can not foster a sense of accomplishment to fuel motivation. On the other hand, goals with unrealistic difficulty can also reduce motivation and lead to abandonment as they can be detrimentally depressing. Therefore, goals should be attainable. It is also very important to renew the goal setting and start another goal after the previous goal is reached. With proper goals, people tend to perform better on goal-related tasks, which is language learning in our case.

\subsection{Incentive Planning}
\label{subsec:incentive}

The formation of boredom is often due to the insufficiency of new incentives. Therefore, designing an effective incentive plan is of great importance in withstanding the repetitiveness in language learning. Proper incentives are needed to power online learning platform users through this emotional state. Incentives can be provided both directly and indirectly. Direct incentives in language learning involve bringing in novelty and changes to help break free from repetitive learning patterns and rekindle interest. Possible ways include changing the learning materials, learning modes, etc. Indirect incentives are supplemental rewards set as a motivational device to increase self-recognition. They can provide important feedback on short-term achievements, which in turn can boost perseverance in achieving long-term goals. Incentives, both direct and indirect, are indispensable in long-term language learning, especially before the daily learning behavior becomes habitual. Language learners should thoughtfully incorporate incentive planning into their language learning journey.

\subsection{Light Starting}
\label{subsec:light}

Users of online learning platforms should recognize that in the beginning, the overarching task lies in cultivating a new daily habit, and knowledge acquisition at this point is of secondary importance. Habit formation should begin with a manageable light starting. However, individuals often overestimate their willpower and pursue overly ambitious strategies aimed at rapid progress. This frequently results in failure to sustain consistent engagement. We should follow the principle of ``less is more'' and take one small step a day. This is particularly relevant for passion-driven learners, who are frequently inspired by external stimuli such as speeches, role models, or literature. With strong feelings towards mastering a language, such individuals are more likely to overestimate themselves and start with heavy daily workloads. When they later struggle to meet these demands, their passion can easily be dampened by frustration, which will cause the abandonment of language learning. Attempting to learn more each day may conflict with other daily responsibilities, further reducing motivation. In contrast, a light daily task promotes sustainability in language learning, easing psychological burden. Achieving a small step on a regular basis can also build a sense of accomplishment, which will gradually increase motivation and maintain passion for language learning. Most importantly, a light starting is conducive to transforming language learning into a daily habit for long-term daily routine.

\subsection{Team Learning}
\label{subsec:team}

Learning in isolation is often suboptimal, particularly in the domain of language acquisition. Humans are inherently social beings, and this social nature extends to learning processes as well. \citet{le-crowd-2017} illustrates the crowd influence on individuals' behavior. Compared to individual learning, team learning has been shown to enhance performance across various dimensions~\citep{vasan-team-2011}. In online learning environments, teams can function as a robust support system for platform users, offering encouragement to beginners, providing motivation, and facilitating collaborative problem-solving, among other benefits. In addition to intra-team cooperation, inter-team competition can also further enhance motivation and cooperation. Moreover, team supervision stands as a desirable accompaniment, pushing users to continue with the long-term journey. Prior to the establishment of strong learning habits, individuals are particularly susceptible to forgetting the daily small steps. In such cases, reminders from the team or team members serve as effective mechanisms in habit formation. Social accountability and shared goals can significantly increase persistence and goal attainment. Therefore, incorporating team dynamics into language learning can lead to better overall learning outcomes.

\section{Usage Instructions}
\label{sec:tips}

\subsection{The Early Stage}
\label{subsec:early}

Approximately the first two months from starting using Shanbay is defined as the early stage. This stage is of critical importance for cultivating perseverance and fostering awareness. \citet{lally-how-2010} reveal that the median time of repetition needed to form a habit is 66 days. Building a new habit requires forming a new routine, but many obstacles can hinder this process. In the first few days right after beginning to use Shanbay, users can easily complete the daily check-in with their passion and ambition to start a new habit. However, the hardest part is to continue after those initial days, and most of the users, whether willingly or unwillingly, give up during this period due to reasons such as insufficient time, forgetfulness, etc.

I suggest that new users begin with the recognition mode of identifying the Chinese meaning of English words on Shanbay Words. These words are recommended to be from a vocabulary list with no more than 300 words and at a level slightly higher than the user's current level. For the daily task at the early stage, new users are suggested to learn 15 new words and review 30 words. It takes about 10 to 15 minutes to finish this task for the check-in, making it easy to accomplish without creating a burden. This task shall remain the same without any extra workload added throughout the early stage, even if the new user finds it effortless to finish. This daily task is set at the minimum level for the purpose of forming a new habit through this stage and preventing early dropping out caused by insufficient time.

There are two handy functions available on Shanbay. The first function is to pair with another user as study buddies. After pairing up, both users can view each other's learning records and status. Study buddies can also send check-in reminders to each other. After checking in as a pair for certain days, a special achievement will be rewarded. The second function is to join a study group. These groups are divided based on users' backgrounds. Users in the same group might share the same profession, school, learning goal, or education degree, etc. A ranking list of the check-in data of each group member is presented for every group, fostering a sense of competition. Both of these two options are for the purpose of team learning, which can be an effective way to overcome forgetfulness. Users may also choose to team up on their own without using these two functions provided by Shanbay.

\subsection{The Middle Stage}
\label{subsec:middle}

Users who follow the methods above and possess a strong ambition to improve themselves can manage to persist through the early stage and enter the middle stage. From the third month to approximately the second year is defined as the middle stage, though the exact time period may vary from person to person. The criterion in differentiating between the middle stage and the late stage is that whether users consider learning with the online platform and checking in as an internalized long-term behavior. At this stage, the habit of check-in is initially formed but still requires reinforcement, as the initial passion for undertaking the course is fading. Therefore, maintaining perseverance at the middle stage calls for new goals and fresh incentives.

I suggest that, at the middle stage, users should sign up for a standard English test and set passing it as a concrete goal. The chosen test should be able to pose reasonable challenges to the user's English proficiency. Despite being controversial with the validity, the results of standard English tests serve as a comparatively reliable reference to the user's language ability. Passing a standard test can improve the user's self-recognition once the user realizes that the efforts made online can produce tangible real-life results. After passing one test, users should choose a more advanced test and continue this cycle to maintain motivation and progress.

For incentive planning at the middle stage, I have three main suggestions. First, switch to Shanbay Reading or Shanbay Speaking and Listening for the daily task. Memorizing English words is an overly simple task in language learning, which is more likely to cause boredom. That is why at the middle stage, users should adopt more complicated but interesting materials for more diversified learning activities, including news reading, listening exercises, and speaking practice. Second, set short-term goals for more regular rewarding. At this stage, users can gradually increase their workload, but such workload should be a short-term goal achievable within two to three months. Third, share the check-in records and learning progress. Users should post their milestone check-in records, such as 100, 200, or 500 days of checking in, on social media. It is also helpful to write down personal experiences and learning tips, and share to the Shanbay community or social media. Likes and comments from others can generate great stimulation for users to persevere in repetitive daily tasks.

\subsection{The Late Stage}
\label{subsec:late}

With about two years of continued learning and checking in on the online platform, this behavior will have been internalized to be an individual's natural daily routine. At this stage, the check-in behavior becomes physiological, just like eating and drinking, and this marks the transition into the late stage. Nonetheless, interruption can still occur and disrupt this natural behavior, just as malfunctions can happen to machines. There are two types of factors in this interruption: internal and external.

The external factors are special events that can be called ``accidents''. These ``accidents'' are capable of changing a person's planned daily routine, taking up the whole day and draining all the energy. With these ``accidents'', the user may not be able to do anything else, including the daily check-in. The user might have check-in reminders sent by Shanbay or the study buddy, but is unable to check or react to the reminding message. Under this circumstance, the missed check-in is not due to a lack of intention, but purely because the user can not spread too thin.

``Accidents'' are unpreventable, but remedies are available as Shanbay offers a supplementary check-in service. It allows users to make up for a missed check-in and maintain a continual record, provided that one of the following two conditions is met. One is to consecutively check in for 21 days, and the other is to pay for Shanbay's VIP membership. These remedies are particularly helpful in cultivating long-term perseverance. According to the Broken Window Theory, to prevent further disorder of absenteeism, effective actions and management strategies should be taken~\citep{juliana-study-2019}. In the context of the check-in service, fixing a missed check-in to preserve a continual record is just the same as fixing a ``broken window''. This serves as a powerful motivation for users to persevere in their learning with the online platform.

For the internal factor, emotion is what leads to the interruption. Many studies show that individuals might find themselves holding an unreasonable disgust towards their daily behavior at certain points. They may feel the urge to escape from or fully abandon this behavior. This emotion is referred to as burnout~\citep{li-conceptualisation-2024}. If this happens, forcing oneself to continue the behavior can backfire, as this can add to the severity of disgust. It should be noted that the emotion of disgust is different from laziness or unwillingness to persevere in learning, as the emotion of disgust only arises at the late stage when the check-in behavior has already become physiological.

As soon as the emotion of disgust is felt, the user should take a break from the daily check-in routine. The purpose of the check-in behavior is to achieve lifelong learning, but when disgust arises, it undermines that purpose. That is why a temporary break is necessary to restore a healthy relationship with the routine. The duration of the break should be from one to two weeks. Breaks longer than two weeks are not recommended as they could result in rebuilding the habit from scratch. During this time, the user should reach out for help, asking a family member or a friend to check in on their behalf to preserve a straight check-in record. Again, a continual check-in record plays a crucially important role in this long-term journey of persevering in language learning.

\section{Conclusion}
\label{sec:concl}
In this paper, I have examined the issues of user abandonment of online English learning platforms with a check-in service. The GILT method and its instructions are proposed as a detailed practical solution to address these issues. Endorsed by other users who have applied the method, the GILT method has been proved to be feasible and efficient in cultivating perseverance in real life. Moreover, with its insights into the online learning behavior, the GILT method is also applicable to other online learning platforms for habit formation.

\section*{Acknowledgment}

Special and heartfelt gratitude goes to my wife Fenmei Zhou, for her understanding and love. Her unwavering support and continuous encouragement enable this research to be possible.


\bibliographystyle{IEEEtranN}
\bibliography{mybib}


\end{CJK}

\end{NoHyper}   
\end{document}